\documentstyle[emulateapj,epsfig]{article}

\begin{document}

\lefthead{Fender et al.}
\righthead{Circularly polarised radio emission from SS 433}
\slugcomment{Accepted for publication in ApJ Letters}

\title{
Discovery of circularly polarised radio emission from SS 433
}

\author{Robert Fender\altaffilmark{1}, 
David Rayner\altaffilmark{2},
Ray Norris\altaffilmark{3},
R.J. Sault\altaffilmark{3},
Guy Pooley\altaffilmark{4}
}

\altaffiltext{1}{Astronomical Institute
`Anton Pannekoek', University of Amsterdam, and Center for High Energy
Astrophysics, Kruislaan 403, 1098 SJ Amsterdam, The Netherlands}
\altaffiltext{2}{School of Physics and Mathematics, University of Tasmania,
GPO Box 252-21, Hobart, TAS 7001, Australia}
\altaffiltext{3}{Australia Telescope National Facility, CSIRO,
P.O. Box 76, Epping, NSW 1710, Australia}
\altaffiltext{4}{Mullard Radio Astronomy Observatory, Cavendish Laboratory,
Madingley Road, Cambridge CB3 0HE, U.K.}

\begin{abstract}

We report the discovery of circularly polarised radio emission from
the radio-jet X-ray binary SS 433 with the Australia Telescope Compact
Array. The flux density spectrum of the circular polarization, clearly
detected at four frequencies between 1 -- 9 GHz, is of the form $V
\propto \nu^{-0.9 \pm 0.1}$. Multiple components in the source and a
lack of very high spatial resolution do not allow a unique determination of
the origin of the circular polarization, nor of the spectrum of
fractional polarization. However, we argue that the emission is likely
to arise in the inner regions of the binary, possibly via
propagation-induced conversion of linear to circular polarization, and
the fractional circular polarization of these regions may be as high
as 10\%. Observations such as these have the potential to investigate
the composition, whether pairs or baryonic, of the ejecta from X-ray binaries.

\end{abstract}


\section{Introduction}

High-velocity synchrotron-emitting jets are commonly observed from
Active Galactic Nuclei (AGN; e.g. Ostrowski et al. 1997; Shields
1999), and X-ray binary systems (XRBs) containing both black holes and
neutron stars (e.g. Hjellming \& Han 1995; Fender 1999 and references
therein). The composition of the jet plasma, whether electron-proton
(e$^-$p$^+$) or electron-positron (e$^-$e$^+$) remains a fundamental
yet unanswered question in nearly all cases.

SS 433 is one of the most celebrated of Galactic objects. The source
is an XRB consisting, most probably, of a mass-losing star in a 13-day
orbit with a stellar-mass black hole or neutron star. The system
produces bright quasi-continuous radio jets which precess with a
period of $\sim162.5$ days (Vermeulen 1989; Vermeulen 1992; Brinkmann
1998). Moving optical emission lines (Margon 1984 and references
therein) indicate a jet velocity of $\beta = v/c = 0.26$, confirmed by
both VLA and VLBI radio observations (Vermeulen 1992 and references
therein). These optical lines, and their X-ray counterparts (Kotani et
al. 1996 and references therein), are the only direct evidence for the
existence of baryonic material (i.e. e$^-$p$^+$) in a jet from any
X-ray binary.

Recent progress towards determining the composition of the plasma in
jets from AGN has been made by the detection and modelling of a
circularly polarised radio component from the quasar 3C 279 (Wardle et
al. 1998). Wilson \& Weiler (1997) argue that radio circular
polarization upper limits for the Crab supernova remnant come close to
determining the positron content of the nebula.  In addition Bower,
Falcke \& Backer (1999) and Sault \& Macquart (1999) have recently
detected circularly polarised radio emission from Sgr A* at the
Galactic Centre.

In this paper we report the detection of circularly polarised radio
emission from SS 433, the first from any XRB, at four radio
frequencies. This observation has the potential to be the benchmark
against which other jets from other XRBs may be compared in an effort
to determine whether they produce e$^-$p$^+$ or e$^-$e$^+$ jets.

\begin{figure*}
\centerline{
\epsfig{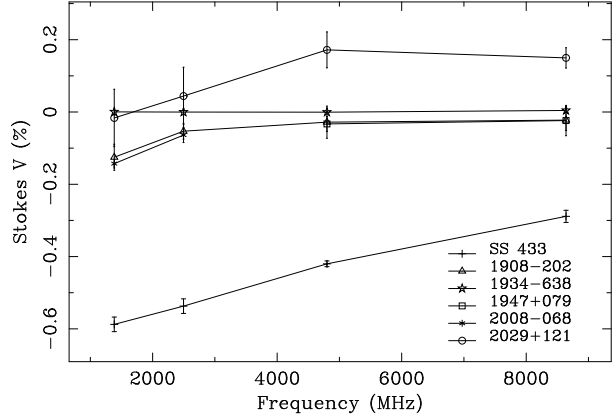}}
\caption{Fractional circular polarization for SS433 and calibrators
for 1999 May 20 observations. The scale is set by assuming PKS
1934-638 has Stokes V=0 at all frequencies (see text).}
\end{figure*}

\begin{figure*}
\centerline{
\epsfig{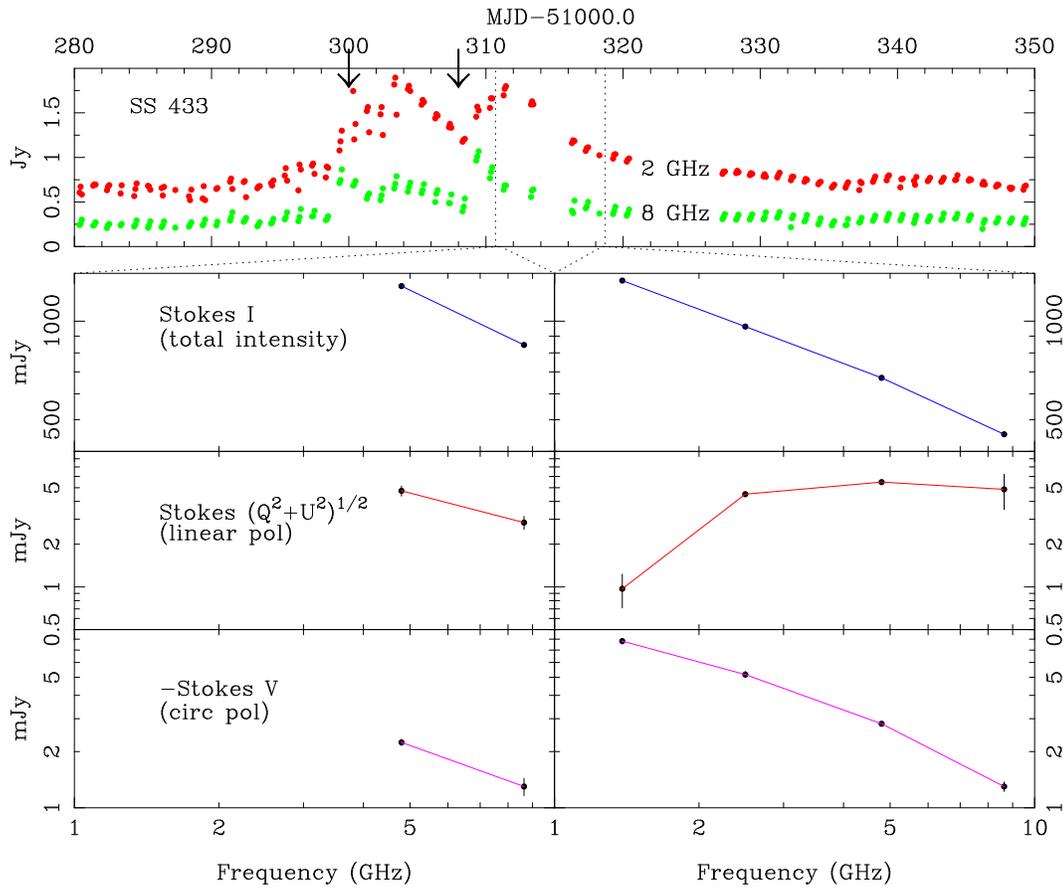}}
\caption{Observations of SS 433 in 1999 May. Top panel illustrates GBI
(quasi-)daily monitoring at 2.3 \& 8.3 GHz; our best estimates of the
ejection dates corresponding to the two major flares are indicated by
arrows. The lower panels indicate mean
flux densities in total intensity, linear and circular polarizations
for the two epochs of ATCA observations.}
\end{figure*}


\section{Observations and results}

\begin{figure*}
\centerline{
\epsfig{file=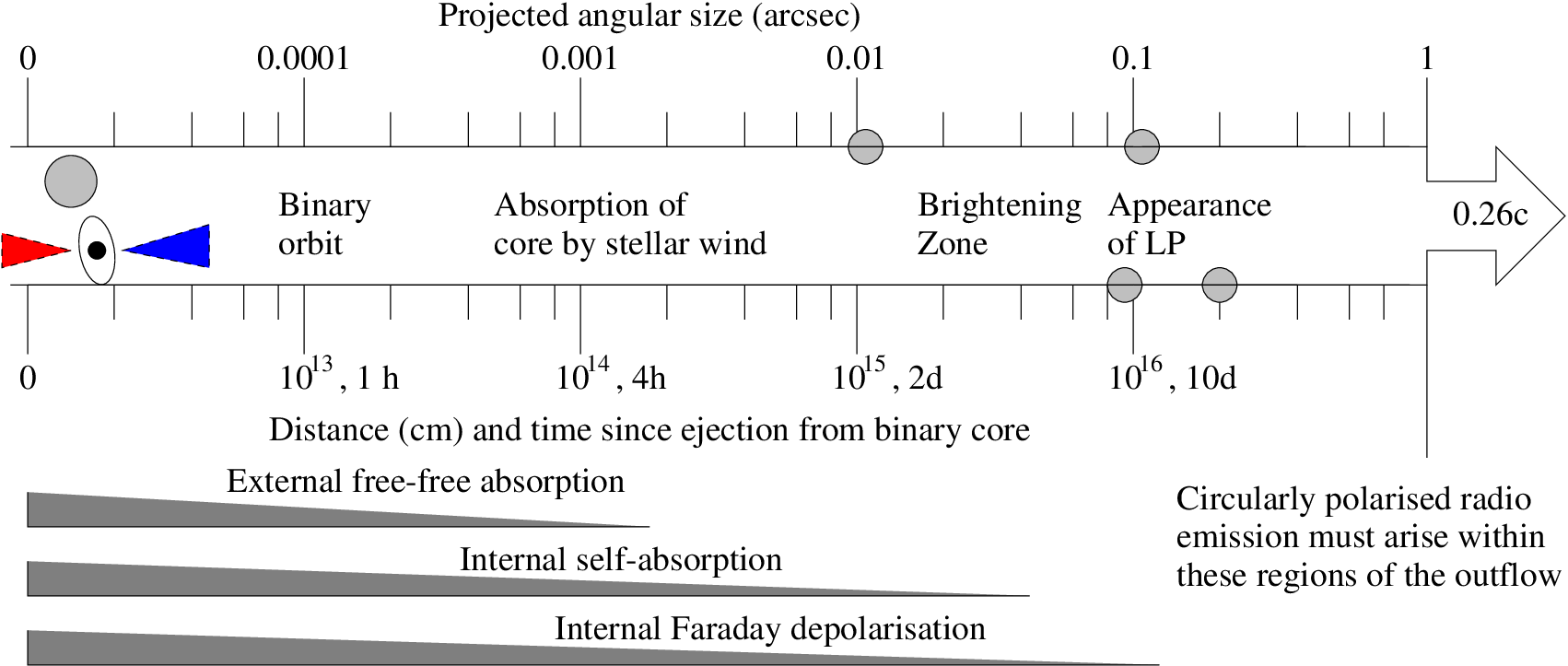,angle=270,width=14cm,clip}}
\caption{
Progression of internal and external absorption and depolarization
effects as ejecta move away from SS 433 
(Hjellming \& Johnston 1981; Vermeulen 1992; Paragi et al. 1999).
The circles at the top and bottom of the schematic indicate our best
estimates of the physical locations of the components responsible for
the radio flares which dominate the total flux density, on 1999 May 10
and May 20 respectively.
}
\end{figure*}

The Australia Telescope Compact Array ({\em ATCA}) consists of six
22~m alt-az antennas near Narrabri, New South Wales (Frater, Brooks,
\& Whiteoak 1992). Each ATCA antenna is equipped with two wide-band
feed horns, and each feed horn is equipped with two pairs of
orthogonal-linear probes. This allows both orthogonal polarizations at
two separate frequencies to be observed with each feed horn at the
same time. Observations were made using the ATCA `continuum-mode',
which gives bandwidths of 128~MHz simultaneously at each of two
frequencies, and four correlation products (XX, YY, XY, YX). The
observations on 1999 May 10 were made centered on 4.80~GHz and
8.64~GHz. On 1999 May 20, observations at 1.38~GHz and 2.50~GHz were
alternated with those at 4.80~GHz and 8.64~GHz, with a cycle time of
$\sim25$~minutes. Calibration sources, selected from the ATCA
Calibrator Source Catalogue (Reynolds 1997), were observed every
cycle. The ATCA primary calibrator PKS~1934--638 was used to calibrate
the bandpass and to set the absolute flux scale (Reynolds 1994).

Data reduction was performed with the \textsc{Miriad} package (Sault,
Teuben, \& Wright 1995).  Prior to gain calibration, the xy-phase
correction measured by a noise-diode system was applied, and the data
were corrected for a small (otherwise unmodelled) field rotation due
to the antenna's pointing model (Kesteven 1997). The main calibration
step involves simultaneously solving for time-dependent complex-gains,
time-independent residual xy-phase, and time-independent polarization
leakages for each feed, as well as the linear and circular
polarization of the calibration source. A good parallactic-angle
coverage is required in order that the leakages and calibrator linear
polarization can be decoupled in the solution process through the
relative rotation of the feeds and parallactic angle (Conway \&
Kronberg 1969).

Circular polarization data requires calibration using the
`strongly-polarized' equations (Sault, Killeen, \& Kesteven 1991),
which include terms in $leakage \times \left({\mbox Stokes}\
Q,U\right)$. To obtain such a solution requires a calibration source
which has a few percent linear polarization (for this experiment
PKS~1908-202) so that there is sufficient signal in the second-order
terms.  The accuracy of the leakage solution was estimated by
repeating the entire calibration procedure for the 1999 May 20
observation, for all four frequencies, with a different calibrator
(either PKS~1947+079 or PKS~2029+121). For SS433, the differences in
the resultant circular polarization are of the same order as the
errors expected from system noise alone; these differences have been
incorporated into our circular polarization error estimates.

All calibrators were imaged in circular polarization as a consistency
check, and the results for 20 May, 1999, are shown in Fig 1. The
leakage calibration of these calibrators was constrained so that the
results gave zero circular polarization for PKS 1934-638 (an absolute
circular polarization calibrator is needed for the interferometer --
see Sault, Hamaker \& Bregman 1996).  A bias in the observed circular
polarization of compact steep spectrum sources observed at the ATCA
(Rayner et al. 1999) suggests that PKS~1934--638 is in fact
circularly-polarized with $V/I\sim+2.5\times10^{-4} \pm
0.5\times10^{-4}$ at 4.8~GHz.  An error of this order in the absolute
circular polariation flux-scale is much smaller than the observed
circular polarization of SS433, and does not affect the conclusions of
this paper.  The sign convention of V follows the IAU convention
(Transactions of the IAU, vol 15B, (1973) 166), which conforms to the
IEEE definition (1969, Standard Definitions of terms for radio wave
propogation, IEEE Trans AP-17, 270).

SS 433 was slightly resolved (i.e. $\ga 1$ arcsec) in the E-W
direction in both total intensity and linear polarization, consistent
with the observations of Hjellming \& Johnston (1981).  However the
source is unresolved in Stokes V, consistent with all the circularly
polarised flux arising in an unresolved point source.

On 1999 May 10 we also observed the radio-jet X-ray binary GRS
1915+105; at the time the source had a flat spectrum between 4.80 --
8.64 GHz, at a flux density of $\sim 20$ mJy. We can place $3\sigma$
limits of 1.2\% on both the fractional linear and circular
polarization of the emission from this source.

In addition to our ATCA observations, we also utilise data from the
Green Bank Interferometer (GBI) variable source monitoring
program. These data provide a long-term view of the state of activity
of the radio source with daily flux density measurements at 2.3 and
8.3 GHz (see e.g. Waltman et al. 1991).

\section{Evolution of the propagating ejecta}

Both the GBI monitoring data and our ATCA observations are presented
in Figure 2. Note that the lower two panels show the flux density
measured in linear and circular polarization respectively (i.e.
not the {\em fractional} spectra). A best fit to the observed flux density
spectrum of the circular polarization has a spectral index of 
$\alpha_V = -0.9 \pm 0.1$ (where $V \propto \nu^{\alpha_V}$).
Between the two epochs it is clear that the total radio
flux dropped by $\geq 40$\% and was dominated by the decaying stages
of the two major flare events which peaked, at 2 GHz, on $\sim$ MJD
51304 and $\sim$ MJD 51312 respectively. Over the same period the
linearly polarised flux may have increased slightly, and the
circularly polarised flux appears to have remained constant.

Our current understanding of the evolution of external and internal
opacity effects in the outflow from SS 433 is summarised in Fig 3
(based upon Hjellming \& Johnston 1981, Vermeulen 1992 and Paragi et
al. 1999). Also indicated are our best estimates of the physical
locations of the components corresponding to the two flare events, at
the times of our observations on 1999 May 10 and May 20.
We can presume that on our first epoch we only observed linear
polarization from the first ejection, which contributed $\leq 1/3$
of the total flux density at this time. By our second epoch of
observations both components contributed to the linearly polarised
flux. Qualitatively this can explain the increase in the linearly
polarised flux density between the two epochs, even though the total
flux density decreased. Qualitatively similar behaviour is seen in the
evolution of linear polarization in ejections from GRS 1915+105
(Fender et al. 1999) where the core also remains persistently
(linearly) depolarised.

\section{The origin of the circular polarization}

Circular polarization may be produced in a synchrotron-emitting plasma
either directly as a result of the synchrotron process or via
conversion of linear to circular polarization (Kennett \& Melrose 1998
and references therein). Below we briefly discuss the possible
interpretations of the circularly polarised radio emission from SS 433
in the context of these models.

\subsection{Intrinsically circularly polarised synchrotron emission}

An electron of Lorentz factor $\gamma$ can be considered to radiate
synchrotron emission primarily at a frequency $\nu = 4.2 B_{\perp}
\gamma^2$ MHz, where $B_{\perp}$ is the component of the magnetic
field perpendicular to the line of sight, measured in Gauss.  The
fractional circular polarization $m_c$ (= Stokes $|V|/I$) produced
intrinsically by synchrotron radiation is of order $1/\gamma$ (Legg \&
Westfold 1968), and hence the observed circular polarization spectrum
should follow the relation $m_c \propto \nu^{-1/2}$. For an observed
$m_c \sim 0.005$ at 8640 MHz, we can estimate a magnetic field
strength of $\sim 50$ mG (corresponding to $\gamma \sim 230$).

This estimate is in order-of-magnitude agreement with magnetic field
estimates for major ejections from SS 433 and other X-ray binaries
(e.g. Hjellming \& Han 1995, Fender et al. 1999). Note in addition
that `Faraday depolarization', which can severely reduce the observed
linear polarization, will have no effect on circular polarization. It
is therefore possible to have an optically thin synchrotron source
which displays a large ratio of circular to linear polarization, as is
observed in this case. However, reversals in the line-of-sight
component of the magnetic field which are likely to occur within the
source will require significantly higher magnetic fields.  Similarly,
a significant e$^+$ population within the source will also cause a
reduction in the observed $V$, and require a significantly higher
magnetic field.  In addition, the $\sim$ constant circularly polarised
flux density at 8640 MHz during a drop by $\geq 40$\% in the total
flux density does not seem compatible with an origin for the circular
polarization in the optically thin ejecta which correspond to the two
major flares.

\subsection{Propagation-induced circular polarization}

Linearly polarised radiation can be converted to circularly polarised
radiation during propagation through a plasma with elliptical (or
linear) propagation modes (Pacholczyk 1973; Kennett \& Melrose 1998
and references therein). In the event of the admixture of a small
amount of relativistic plasma to a thermal plasma, propagation modes
through the plasma will become slightly elliptical, and a spectrum of
the form $m_c \propto \nu^{-1}$ is predicted (Pacholczyk 1973). In the
event of plasma which is dominated by highly relativistic particles,
the propagation modes may approach linear, and a much steeper spectrum
of the form $m_c \propto \nu^{-3}$ is predicted (Kennett \& Melrose
1998). 

Wardle et al. (1998) argued that the circular polarization observed
from 3C 279 arose because of such propagation-induced
`repolarization'. They concluded that the low-energy spectrum of the
relativistic particles must extend to $\gamma << 100$, and therefore
the jet must be composed of an e$^+$e$^-$ plasma (if there were
protons accompanying each emitting electron the kinetic energy of the
jet would be several orders of magnitude greater than that which is
seen to be dissipated at the head of the jet). We note that analogous
considerations may be also applicable to SS 433, where X-ray and radio
hotspots are observed within the W 50 radio nebula, presumably at the
site of the jet : ISM interaction. Indeed, the kinetic energy in the
jets of SS 433, if they are composed of a e$^-$p$^+$ plasma, is $\geq
10^{40}$ erg s$^{-1}$ (Brinkmann et al. 1988), which is much greater
than that which is directly observed to be dissipated on larger scales
within W50 (this is one of the arguments against an e$^-$p$^+$ plasma
which is put forward by Kundt 1998).

\subsection{Alternative mechanisms ?}

Given the complexity of the SS 433 system and the high densities and
magnetic field strengths likely to be present near to the base of the
jet, alternative origins for the circularly polarised emission cannot
be ruled out. These include gyrosynchrotron emission from low-energy
electrons and cyclotron maser emission (Dulk 1985). We note that if
the circularly polarised emission is associated with a region on the
scale of one of the binary components (e.g. $<10^{12}$ cm) then the
brightness temperature is $\geq 10^{10}$ K at 8.64 GHz, and $\geq
10^{12}$ K at 1.4 GHz.

\section{Discussion}

It is of great importance to determine in which of the various
emitting regions the circularly polarised flux density originated, in
order to determine both the spectrum and relative strength of the
emission.  Our ATCA observations rule out an extension with the
optically thin jets on scales of $\geq 1$ arcsec, and the lack of
correlated variability in Stokes I and V argues against an association
with the two major ejection events. The underlying radio spectrum of
SS 433 is typically around 250 mJy at 8 GHz, with an optically thin
spectral index of about $-0.7$, corresponding to a quasi-continuous
flow of matter into the jets.  If associated with this component then
the spectrum of the relative circular polarization could be as flat as
the $m_c \propto \nu^{-1/2}$ predicted for intrinsic synchrotron
emission.  However, if associated with this component, then why not
with the two flares, which are presumably just enhancements of the
same flow ? Alternatively, Paragi et al. (1999) show that the
innermost regions ($\leq 50$ mas) of the jets have a flat/inverted
spectrum between 1 -- 15 GHz. The core region has a peak flux density
typically of $\leq 100$ mJy on VLBI scales. If associated with this
region the fractional circular polarization may be as high as 10\% and
the spectrum may steepen to the $m_c \propto \nu^{-1}$ predicted for a
mildly relativistic plasma (it seems unlikely that the $m_c \propto
\nu^{-3}$ spectrum can be recovered, unless the emission arises right
in the binary core of the system, which has the most inverted radio
spectrum). If associated with the inner regions, this implies a very
large ratio of circular to linear polarizations, as found for Sgr A*
by Bower et al. (1999).

Further precessional-phase-resolved monitoring and spatial resolution
of the regions responsible for the circular polarization are essential
to further investigate this discovery. If, as seems likely, the
circular polarization is associated with the synchrotron emitting
ejecta, comparison with circular polarization measurements of other
X-ray binaries has the potential to reveal, finally, the composition
of the relativistic plasmas.

\acknowledgements

RPF would like to acknowledge useful discussions with Ralph Spencer
and Al Stirling, and to thank Mark Walker for the original suggestion
to look for circular polarisation from X-ray binaries.  The Australia
Telescope is funded by the Commonwealth of Australia for operation as
a National Facility managed by CSIRO. RPF was funded during the period
of this research EC Marie Curie Fellowship ERBFMBICT 972436.

\end{document}